\documentstyle[preprint,aps,prb]{revtex}
\begin{document}
\tighten
\title{Nonlinear transport in inelastic Maxwell mixtures under simple shear flow}
\author{Vicente Garz\'{o}\footnote[1]{Electronic address: vicenteg@unex.es}}
\address{Departamento de F\'{\i}sica, Universidad de Extremadura, E-06071 \\
Badajoz, Spain}

\date{\today}
\maketitle

\begin{abstract}
The Boltzmann equation for inelastic Maxwell models is used to analyze nonlinear transport in a granular binary mixture in the steady simple shear flow. Two different transport processes are studied. First, the rheological properties (shear and normal stresses) are obtained by solving exactly the velocity moment equations. Second, the diffusion tensor of  impurities immersed in a sheared inelastic Maxwell gas is explicitly determined from a perturbation solution through first order in the concentration gradient. The corresponding reference state of this expansion corresponds to the solution derived in the (pure) shear flow problem.  All these transport coefficients are given in terms of the restitution coefficients and the parameters of the mixture (ratios of masses, concentration, and sizes).  The results are compared with those obtained analytically for inelastic hard spheres in the first Sonine approximation and by means of Monte Carlo simulations. The comparison between the results obtained for both interaction models shows a good agreement  over a wide range values of the parameter space.   
  

Running title:  Inelastic Maxwell mixtures

\end{abstract}

\draft


\bigskip \narrowtext

\newpage

\section{Introduction}
\label{sec1}

Granular media under rapid flow conditions are usually modeled by a fluid 
of hard spheres with inelastic collisions. In the simplest version, the 
grains are taken to be smooth so that the inelasticity is only characterized 
through a constant coefficient of normal restitution.  For a low-density 
gas, the Boltzmann equation has been conveniently generalized \cite{BDS97}   
to account for the inelasticity of binary collisions and the Navier-Stokes 
transport coefficients have been obtained in terms of the restitution 
coefficient.\cite{BDKS98,GD99,GM02,GD02} As in the case of elastic 
collisions, these transport coefficients verify a set of coupled linear 
integral equations which are solved approximately by using the leading terms 
in a Sonine polynomial expansion. The corresponding Sonine  
results are in general in good agreement with selected tests using molecular 
dynamics \cite{LBD02} and Monte Carlo simulations.\cite{GM02,BMC99,MG03} 
However, beyond the Navier-Stokes regime (small spatial hydrodynamic 
gradients), it is quite intricate to get explicit results for inelastic hard 
spheres (IHS) and one has to resort to alternative approaches.

One of the main mathematical difficulties in solving the Boltzmann equation 
for hard spheres (even in the elastic case) comes from the form of the 
collision rate, which is proportional to the magnitude of the relative 
velocity of the two colliding spheres. In the case of elastic fluids, a 
possible way to overcome this problem is to assume that the particles 
interact via the repulsive Maxwell potential (inversely proportional to the 
fourth power of the distance). For this interaction model, the collision 
rate is independent of the relative velocity and this allows for a number of nice 
mathematical properties of the Boltzmann collision operator.\cite{E81} 
Thanks to this simplification, nonlinear transport properties can be exactly 
obtained \cite{SG95} from the Boltzmann equation for Maxwell elastic 
molecules and, when properly reduced, they exhibit a good agreement with 
results obtained for other interaction models.\cite{Pepe} In the context of 
inelastic gases, the Boltzmann equation for inelastic 
Maxwell models (IMM) was also introduced about two years ago.\cite{NK00,BCG00,CCG00} 
The IMM share with elastic Maxwell molecules the property that the collision 
rate is velocity independent but their collision rules are the same as for 
IHS. As a consequence, these IMM's do not describe real particles since 
they do not interact according to a given potential law. For this reason 
the model is usually referred to as pseudo-Maxwellian model.\cite{BCG00} 
Nevertheless, the IMM keeps the qualitatively correct structure and 
properties of the nonlinear macroscopic equations and obey 
Haff's law.\cite{EB02}

The Boltzmann equation for IMM has received a great attention in the last 
few years, especially in the study of overpopulated high energy 
tails in homogeneous states.\cite{EB02,BMP02,KN02,EB02bis} 
The existence of high energy tails of the Boltzmann equation is common for IHS and IMM, although this general qualitative agreement fails at a quantitative level. For inhomogeneous states,  the Chapman-Enskog method has been recently applied to the Boltzmann equation to get the Navier-Stokes transport coefficients.\cite{S02,H02} The comparison with the results derived for the IHS model\cite{BDKS98} shows in general significant discrepancies in the dependence of the transport coefficients on the restitution coefficient. The IMM has been also used to get the shearing stress of a granular material under shear flow.\cite{C01}All the above results refer to monocomponent gases. Much less is known for inelastic Maxwell mixtures.  In the case of multicomponent gases, Marconi and Puglisi\cite{MP02} have analyzed energy nonequipartition in the free cooling and driven states for a one-dimensional system, while Ben-Naim and Krapivsky \cite{NK02} have studied velocity statistics of an impurity inmersed in a uniform granular fluid. However, to the best of my knowledge, no previous study on transport properties in inelastic Maxwell mixtures has been carried out.

The aim of this paper is to analyze nonlinear transport in a binary mixture in the framework of the Boltzmann equation for IMM. Two different transport processes will be studied. First, the rheological properties (shear stress and normal stress differences) of a granular binary mixture subjected to the simple shear flow will be explicitly determined.  Second, the elements of the tracer diffusion tensor characterizing the diffusion of impurities in a granular gas under shear flow will be obtained in the first order of the concentration gradient. In both situations, given the coupling between dissipation and the shear rate, the mixture is far away from equilibrium and the Navier-Stokes constitutive equations \cite{BDKS98,S02,H02} do not apply. This makes  the analytical treatment of both nonequilibrium situations difficult. An additional complication arises from the large number of parameters governing the dynamics of the system, including three independent restitution coefficients.  However, the tractability of the Boltzmann collision integrals for IMM allows one to determine all the above nonlinear transport properties by solving exactly the hierarchy of moment equations. The results derived for IMM are then compared with those recently obtained for IHS  in the leading Sonine approximation \cite{MG02,G02} and by means of Monte Carlo simulations. \cite{MG02,MG02bis} Since the strength of the shear rate is arbitrary the test of the capability of  IMM to capture the relevant behavior of IHS is quite stringent. The comparison performed here shows that in general the agreement between both interaction models is quite good, even for strong dissipation.   

The plan of the paper is as follows. In Sec.\ \ref{sec2} the Boltzmann equation for IMM and the macroscopic conservation laws are introduced. Section \ref{sec3} deals with the simple shear flow problem for inelastic Maxwell mixtures. The elements of the pressure tensor are exactly obtained in terms of the restitution coefficients and the parameters of the mixture (masses, sizes, and concentrations). In particular, the results show that both species have different temperatures so that  the energy is not equally distributed between both species (breakdown of energy equipartition). The diffusion tensor of impurities in a Maxwellian sheared fluid  is obtained in Sec.\ \ref{sec4}. The paper is closed in Sec.\ \ref{sec5} with some concluding remarks.

\section{The Boltzmann equation and the inelastic Maxwell model}
\label{sec2}

Let us consider a binary mixture of inelastic Maxwell gases at low density.  In the simplest version, 
the Boltzmann equation for IMM \cite{NK00,EB02,EB02bis,S02,NK02} can be 
obtained from the Boltzmann equation for IHS by replacing the  
rate for collisions between particles of species $r$ and $s$ by an average  velocity-independent 
collision rate, which is proportional to the square root of the temperature.  This means that a random pair of colliding particles undergo inelastic collisions with a random impact direction. With this simplification, the set of nonlinear Boltzmann kinetic equations become
\begin{equation}
\label{2.1}
\left(\frac{\partial}{\partial t}+{\bf v}_1\cdot \nabla \right)f_{r}
({\bf r},{\bf v}_1;t)
=\sum_{s}J_{rs}\left[{\bf v}_{1}|f_{r}(t),f_{s}(t)\right] \;,
\end{equation}
where $f_r({\bf r},{\bf v}_1;t)$ is the one-particle distribution function of species $r$ ($r=1,2$) and  
the Boltzmann collision operator $J_{rs}\left[{\bf v}_{1}|f_{r},f_{s}\right]$ describing the scattering of 
pairs of particles is 
\begin{equation}
J_{rs}\left[{\bf v}_{1}|f_{r},f_{s}\right]  =\frac{w_{rs}}{n_s\Omega_d}
\int d{\bf v}_{2}\int d\widehat{\bbox {\sigma }}\left[ \alpha_{rs}^{-2}f_{r}({\bf r},{\bf v}_{1}',t)f_{s}(
{\bf r},{\bf v}_{2}',t)-f_{r}({\bf r},{\bf v}_{1},t)f_{s}({\bf r},{\bf v}_{2},t)\right] 
\;. 
\label{2.2}
\end{equation}
Here, $n_s$ is the number density of species $s$, $w_{rs}$ is an effective collision 
frequency (to be chosen later) for collisions  of type $r-s$,  $\Omega_d=2\pi^{d/2}/\Gamma(d/2)$ is the total solid angle in $d$ dimensions, and $\alpha_{rs}\leq 1$ refers to the constant restitution coefficient for collisions between particles of species $r$ with $s$.   In 
addition, the primes on the velocities denote the initial values $\{{\bf v}_{1}^{\prime},
{\bf v}_{2}^{\prime}\}$ that lead to $\{{\bf v}_{1},{\bf v}_{2}\}$
following a binary collision: 
\begin{equation}
\label{2.3}
{\bf v}_{1}^{\prime }={\bf v}_{1}-\mu_{sr}\left( 1+\alpha_{rs}    
^{-1}\right)(\widehat{\bbox {\sigma}}\cdot {\bf g}_{12})\widehat{\bbox 
{\sigma}},
\quad {\bf v}_{2}^{\prime}={\bf v}_{2}+\mu_{rs}\left( 
1+\alpha_{rs}^{-1}\right) (\widehat{\bbox {\sigma}}\cdot {\bf 
g}_{12})\widehat{\bbox{\sigma}}\;,  
\end{equation}
where ${\bf g}_{12}={\bf v}_1-{\bf v}_2$ is the relative velocity of the colliding pair, 
$\widehat{\bbox {\sigma}}$ is a unit vector directed along the centers of the two colliding 
spheres, and 
\begin{equation}
\label{2.3bis}
\mu_{rs}=\frac{m_r}{m_r+m_s}.
\end{equation}
There is another more refined version of the inelastic Maxwell model \cite{BCG00,CCG00,BC02} where the collision rate  has the same dependence on the scalar product ($ \widehat{\bbox {\sigma }}\cdot {\bf g}_{12}$) as in the case of hard spheres.  The corresponding Boltzmann equation can be proved to be equivalent to Eq.\ (\ref{2.1}) except that the collision laws (\ref{2.3}) must be changed.  
However, both versions of IMM lead to  similar results in problems as delicate as the high energy tails.\cite{EB02}   Therefore, for the sake of simplicity, here I will consider the version given by the collision rules (\ref{2.3}). As will be shown later, this model compares quite well with the results obtained from the IHS model.

At a hydrodynamic level,  the relevant quantities in a binary mixture are the number densities $n_r$, the flow velocity  ${\bf u}$, and the ``granular'' temperature $T$. They are defined in terms 
of moments of the distribution $f_r$ as
\begin{equation}
\label{2.4}
n_r=\int d{\bf v} f_r({\bf v}),  \quad 
\rho{\bf u}=\sum_r\rho_r{\bf u}_r=\sum_r\int d{\bf v}m_r{\bf v}f_r({\bf 
v}),
\end{equation} 
\begin{equation}
\label{2.5}
nT=\sum_rn_rT_r=\sum_r\int d{\bf v}\frac{m_r}{d}V^2f_r({\bf v}),
\end{equation} 
where $\rho_r=m_rn_r$, $n=n_1+n_2$ is the total number density, $\rho=\rho_1+\rho_2$ is the 
total mass density, and ${\bf V}={\bf v}-{\bf u}$ is the peculiar 
velocity. Equations (\ref{2.4}) and (\ref{2.5}) also define the flow 
velocity ${\bf u}_r$ and the partial temperature $T_r$ of species $r$, the latter measuring  
the mean kinetic energy of species $r$. 

The collision frequencies $w_{rs}$ can be seen as free parameters in the model. Its dependence on 
the restitution coefficients $\alpha_{rs}$ can be chosen to 
optimize the agreement with  the results obtained from the Boltzmann equation  for IHS. Of course, 
the choice is not unique and may depend on the property of interest.

The collision operators conserve the particle number of each species and the 
total momentum, but the total energy is not conserved. This implies that 
\begin{equation}
\label{2.6}
\sum_{r,s}\int d{\bf v}\case{1}{2}m_{r}V^{2}J_{rs}
[{\bf v}|f_{r},f_{s}]=-\case{d}{2}nT\zeta \;,  
\end{equation}
where $\zeta$ is identified as the ``cooling rate'' due to inelastic
collisions among all species. At a kinetic level, it is also convenient to 
discuss energy transfer in terms of the ``cooling rates'' $\zeta_r$ for the 
partial temperatures $T_r$. They are defined as 
\begin{equation}
\label{2.7}
\zeta_r=\sum_s \zeta_{rs},
\end{equation}
where 
\begin{equation}
\label{2.7bis}
\zeta_{rs}=-\frac{1}{dn_rT_r}\int d{\bf v}m_rV^{2}J_{rs}[{\bf v}|f_{r},f_{s}]\;.
\end{equation}
The total cooling rate $\zeta$ can be expressed in terms of the partial cooling rates $\zeta_r$ as
\begin{equation}
\label{2.8}
\zeta=T^{-1}\sum_rx_rT_r\zeta_r,
\end{equation}
where $x_r=n_r/n$ is the mole fraction of species $r$.

The macroscopic balance equations follow from the Boltzmann equations (\ref{2.1}) by taking 
velocity moments. They are given by  
\begin{equation} 
D_{t}n_{r}+n_{r}\nabla \cdot {\bf u}+\frac{\nabla \cdot {\bf j}_{r}}{m_{r}} 
=0\;,  \label{2.9} 
\end{equation} 
\begin{equation} 
D_{t}{\bf u}+\rho ^{-1}\nabla {\sf P}=0\;,  \label{2.10} 
\end{equation} 
\begin{equation} 
D_{t}T-\frac{T}{n}\sum_{r}\frac{\nabla \cdot {\bf j}_{r}}{m_{r}}+\frac{2}{dn} 
\left( \nabla \cdot {\bf q}+{\sf P}:\nabla {\bf u}\right) 
=-\zeta T\;. \label{2.11} 
\end{equation} 
In the above equations, $D_{t}=\partial _{t}+{\bf u}\cdot \nabla $ is the 
material derivative,  
\begin{equation} 
{\bf j}_{r}=m_{r}\int d{\bf v}\,{\bf V}\,f_{r}({\bf v})
\label{2.12} 
\end{equation} 
is the mass flux for species $r$ relative to the local flow,  
\begin{equation} 
{\sf P}=\sum_{r}\,\int d{\bf v}\,m_{r}{\bf V}{\bf V}\,f_{r}({\bf  v})  
\label{2.13} 
\end{equation} 
is the total pressure tensor, and  
\begin{equation} 
{\bf q}=\sum_{r}\,\int d{\bf v}\,\case{1}{2}m_{r}V^{2}{\bf V} 
\,f_{r}({\bf v})  
\label{2.14} 
\end{equation} 
is the total heat flux.

The main advantage of using IMM is that a velocity moment of 
order $k$ of the Boltzmann collision operator only involves moments of order less than or equal 
to $k$.\cite{SG95,BC02} This allows one to determine the Boltzmann collision moments without the explicit knowledge of the velocity distribution function.  In general, beyond the linear hydrodynamic regime (Navier-Stokes order), the above property is not sufficient to exactly solve the hierarchy of moment equations due to the free-streaming 
term of the Boltzmann equation. Nevertheless, there exist some particular situations (such as the 
simple shear flow problem) for which the above hierarchy can be recursively solved.  The first few moments of the Boltzmann collision operator $J_{rs}[f_r,f_s]$ have been evaluated in 
Appendix \ref{appA}. They can be written as
\begin{equation}
\label{2.15}
\int d{\bf v}m_r{\bf V} J_{rs}\left[ f_{r},f_{s}\right] = -\frac{w _{rs}}{\rho_sd}\mu_{sr}(1+\alpha _{rs})
\left(\rho_s{\bf j}_r-\rho_r{\bf j}_s\right),
\end{equation}
\begin{eqnarray}
\label{2.16}
\int d{\bf v} m_r {\bf V} {\bf V}J_{rs}[f_r,f_s]&=&
-\frac{w _{rs}}{\rho_sd}\mu_{sr}(1+\alpha _{rs})\left\{2\rho_s{\sf P}_r-\left(
{\bf j}_r{\bf j}_s+{\bf j}_s{\bf j}_r\right)\right. \nonumber\\
& &-\frac{2}{d+2}\mu_{sr}(1+\alpha _{rs})\left[\rho_s{\sf P}_r+\rho_r{\sf P}_s-
\left({\bf j}_r{\bf j}_s+{\bf j}_s{\bf j}_r\right)\right.\nonumber\\
& & \left.\left.
 +\left[\frac{d}{2}\left(\rho_rp_s+\rho_sp_r\right)-{\bf j}_r\cdot {\bf j}_s\right]\openone
\right]\right\},
\end{eqnarray}
where 
\begin{equation}
\label{3.4}
{\sf P}_r=\int d{\bf v}\,m_{r}{\bf V}{\bf V}\,f_{r},   
\end{equation}
and $p_r=n_rT_r=\text{tr}{\sf P}_{r}/d$ is the partial pressure of species $r$. 
The relationship between the collision frequency $w_{rs}$ and 
the cooling rate $\zeta_{rs}$ can be obtained by taking the trace in Eq.\ (\ref{2.16}):
\begin{equation}
\label{2.17}
\zeta_{rs}=\frac{2w _{rs}}{d}\mu_{sr}(1+\alpha _{rs})\left[1-\frac{\mu_{sr}}{2}(1+\alpha_{rs})
\frac{\theta_r+\theta_s}{\theta_s}+\frac{\mu_{sr}(1+\alpha _{rs})-1}{d\rho_sp_r}
{\bf j}_r\cdot {\bf j}_s\right],
\end{equation}
where 
\begin{equation}
\label{2.17bis}
\theta_1=\frac{1+x_1(\gamma-1)}{\mu_{21}\gamma}, \quad 
\theta_2=\frac{1+x_1(\gamma-1)}{\mu_{12}},
\end{equation}
and $\gamma\equiv T_1/T_2$ is the temperature ratio.

\section{Rheological properties under simple shear flow}
\label{sec3}

As said in the Introduction, one of the main objectives of the present paper is 
to evaluate the 
rheological properties of an inelastic Maxwell binary mixture subjected to the simple 
shear flow. At a macroscopic level, this state is characterized by 
a constant linear velocity profile ${\bf u}={\bf u}_1={\bf u}_2={\sf a}\cdot {\bf r}$, 
where the elements of the tensor ${\sf a}$ are 
$a_{ij}=a\delta_{ix}\delta_{j y}$, $a$ being the constant shear rate. 
In addition, the partial densities $n_i$ and the granular temperature $T$
are uniform, while the mass and heat fluxes vanish by symmetry reasons (${\bf j}_r={\bf q}={\bf 0}$). 
As a consequence, the (uniform) pressure tensor ${\sf P}$ is the only nonzero flux in the problem. 
On the other hand, according to the energy balance equation (\ref{2.11}), the temporal variation of the granular temperature arises 
from the balance of two opposite effects: viscous heating and collisional cooling. 
When both mechanisms cancel each other, a steady state is achieved and the temperature remains constant. In that case, the shear stress $P_{xy}$ and the cooling rate $\zeta$ are related by
\begin{equation}
\label{3.1}
aP_{xy}=-\frac{d}{2}\zeta p,
\end{equation}
where $p=\text{tr}{\sf P}/d=nT$ is the pressure.  Our goal here is to determine the pressure tensor ${\sf P}$ in this steady state.

The simple shear flow becomes spatially uniform when one refers the 
velocities of the particles to a frame moving with the flow velocity ${\bf 
u}$: $f_s\left({\bf r},{\bf v}\right)\rightarrow f_s({\bf V})$. 
Consequently, the corresponding steady Boltzmann equations (\ref{2.1}) read  
\begin{equation}
\label{3.2}
-a_{ij} V_{j}\frac{\partial}{\partial 
V_{i}}f_1=J_{11}[f_1,f_1]+J_{12}[f_1,f_2]\;,
\end{equation}
\begin{equation}
\label{3.3}
-a_{ij} V_{j}\frac{\partial}{\partial 
V_{i}}f_2=J_{22}[f_2,f_2]+J_{21}[f_2,f_1]\;.
\end{equation}
The total pressure tensor is ${\sf P}={\sf P}_1+{\sf P}_2$, where 
the partial pressure tensors ${\sf P}_r$ ($r=1,2$) are defined by Eq.\ (\ref{3.4}). 
The elements of these tensors can be obtained by multiplying the Boltzmann equations (\ref{3.2}) and (\ref{3.3}) by $m_r{\bf V}{\bf V}$ and integrating over ${\bf V}$. The result is
\begin{equation}
\label{3.5}
a_{ik}P_{1,kj}+a_{jk}P_{1,ki}+B_{11}P_{1,ij}+B_{12}P_{2,ij}=\left(A_{11}p_{1}+A_{12}p_2\right)\delta_{ij},
\end{equation}
where use has been made of Eq.\ (\ref{2.16}) (with ${\bf j}_r={\bf 0}$) and I have introduced the coefficients
\begin{equation}
\label{3.6}
A_{11}=\frac{w_{11}}{2(d+2)}(1+\alpha_{11})^2+\frac{w_{12}}{d+2}\mu_{21}^2(1+\alpha_{12})^2,
\end{equation}
\begin{equation}
\label{3.7}
A_{12}=\frac{w_{12}}{d+2}\frac{\rho_1}{\rho_2}\mu_{21}^2(1+\alpha_{12})^2,
\end{equation}
\begin{equation}
\label{3.8}
B_{11}=\frac{w_{11}}{d(d+2)}(1+\alpha_{11})(d+1-\alpha_{11})+\frac{2w_{12}}{d(d+2)}\mu_{21}(1+\alpha_{12})
\left[d+2-\mu_{21}(1+\alpha_{12})\right],
\end{equation}
\begin{equation}
\label{3.9}
B_{12}=-\frac{2}{d}A_{12}.
\end{equation}
A similar equation can be obtained for ${\sf P}_2$, by just making the changes $1\leftrightarrow 2$. From Eq.\ (\ref{3.5}), in particular, one gets
\begin{equation}
\label{3.9bis}
aP_{1,xy}=-\frac{d}{2}p_1\zeta_1.
\end{equation}

In order to have explicit expressions one still needs to fix the parameters $w_{rs}$. The most natural choice to find good agreement with the IHS results is to adjust the cooling rates $\zeta_{rs}$ for IMM, Eq.\ (\ref{2.17}), to be the same 
as those obtained for IHS. Given that the cooling rates for IHS are not exaclty known in the simple shear flow problem, I will take for $\zeta_{rs}$ the cooling rate of IHS evaluated at the first Sonine 
approximation: \cite{MG02}
\begin{equation}
\label{3.13}
\zeta_{rs}^{\text{IHS}} \to  \frac{2\Omega_d}{\sqrt{\pi}d}n_s\mu_{sr}\sigma_{sr}^{d-1}v_0\left(
\frac{\theta_r+\theta_s}{\theta_r\theta_s}\right)^{1/2}(1+\alpha_{rs})
\left[1-\frac{\mu_{sr}}{2}(1+\alpha_{rs})
\frac{\theta_r+\theta_s}{\theta_s}\right],
\end{equation}
where $\sigma_{rs}=(\sigma_r+\sigma_s)/2$ and  $v_0=\sqrt{2T(m_1+m_2)/m_1m_2}$ is a thermal velocity defined in tems of the granular temperature of the mixture $T$.  Therefore, according to Eq.\ (\ref{2.17}), the collision frequencies $w_{rs}$ of IMM are given by 
\begin{equation}
\label{3.13bis}
w_{rs}= \frac{\Omega_d}{\sqrt{\pi}}n_s\sigma_{rs}^{d-1}v_0
\left(\frac{\theta_r+\theta_s}{\theta_r\theta_s}\right)^{1/2},
\end{equation}
where use has been made of the fact that ${\bf j}_r={\bf 0}$ in the simple shear flow state. Henceforth, I will present the results of IMM with this choice.

Equation (\ref{3.5}) along with its counterpart for ${\sf P}_2$ constitute a linear system of equations for the total pressure tensor ${\sf P}$. In order to express the solution of this system, it is convenient to introduce dimensionless quantities. As done in a previous study for IHS,\cite{MG02} we reduce the shear rate $a$ with respect to an effective collision frequency 
\begin{equation}
\label{3.9.1}
\nu=\frac{\Omega_d}{2\sqrt{\pi}}n\sigma_{12}^{d-1}v_0.
\end{equation}
Thus, we introduce the reduced cooling rates $\zeta_r^*=\zeta_r/\nu$, the reduced shear rate $a^*=a/\nu$, and the reduced partial pressure tensors ${\sf P}_r^*={\sf P}_r/x_rp$. The reduced total 
pressure tensor is ${\sf P}^*={\sf P}/p=x_1{\sf P}_1^*+x_2{\sf P}_2^*$. It is worthwhile remarking that, for given values of the parameters of the mixture, $a^*$ and ${\sf P}^*$ are {\em only} functions of the restitution coefficients $\alpha_{11}$, $\alpha_{22}$, and $\alpha_{12}$.

In reduced units, the solution to Eq.\ (\ref{3.5}) gives the pressure tensor ${\sf P}^*$ in terms of $a^*$, the temperature ratio $\gamma$, the restitution coefficients and the parameters of the mixture. 
The dependence of $a^*$  on $\alpha_{rs}$ can be obtained from the energy equation (\ref{3.1}) as
\begin{equation}
\label{3.10}
a^*=-\frac{d}{2}\frac{\zeta^*}{P_{xy}^*}=-\frac{d}{2}\frac{x_1\gamma_1\zeta_1^*+x_2\gamma_2\zeta_2^*}
{x_1P_{1,xy}^*+x_2P_{2,xy}^*},
\end{equation}
where I have introduced the temperature ratios $\gamma_1=T_1/T=(\mu_{21}\theta_1)^{-1}$ and 
$\gamma_2=T_2/T=(\mu_{12}\theta_2)^{-1}$.  When the dependence of ${\sf P}_r^*$ and $a^*$ on the temperature ratio $\gamma$ is known, the latter can be obtained by solving Eq.\ (\ref{3.9bis}) (or its counterpart for the species $2$)
\begin{equation}
\label{3.12} 
\gamma=\frac{\zeta_2^*P_{1,xy}^*}{\zeta_1^*P_{2,xy}^*}.
\end{equation}

Once  Eq.\ (\ref{3.12})  is solved, one gets the rheological properties (as measured by the elements of the pressure tensor) of the mixture  in terms of the parameters of the problem.  In the elastic limit ($\alpha_{11}=\alpha_{22}=\alpha_{22}=1$), the cooling rate $\zeta=0$ and the temperature increases in time due to viscous heating ($aP_{xy}\neq 0$).  Consequently, for long times, the reduced shear rate $a^*(t)=a/\nu(t)$ tends to zero and one recovers the equilibrium results, $P_{ij}^*=\delta_{ij}$. Beyond the elastic limit, the reduced pressure tensor has a complex dependence on the restitution coefficients and the parameters of the mixture. Before considering the general case, let us study separately some interesting limit cases.

\subsection{Mechanically equivalent particles}
\label{sec3.1}

This simple case corresponds to $m_1=m_2$, $\sigma_1=\sigma_2$, and $\alpha_{11}=\alpha_{22}=\alpha_{12}\equiv \alpha$. In this limit, Eqs.\  (\ref{3.5}), (\ref{3.10}), and (\ref{3.12}) yield $\gamma=1$, ${\sf P}_1^*={\sf P}_2^*={\sf P}^*$, and the nonzero elements of the (reduced) pressure tensor can be written as
\begin{equation}
\label{3.14}
P_{yy}^*=P_{zz}^*=\cdots=P_{dd}^*=\frac{d}{2}\frac{1+\alpha}{d+1-\alpha},
\end{equation}
\begin{equation}
\label{3.15}
P_{xy}^*=-\sqrt{\frac{d^2(d+2)(1-\alpha^2)}{8(d+1-\alpha)^2}},
\end{equation}
\begin{equation}
\label{3.16}
P_{xx}^*=d-(d-1)P_{yy}^*,
\end{equation}
Expressions (\ref{3.14})--(\ref{3.16}) are independent of the choice of the collision frequency $w_{rs}\equiv w$.

Recently, Cercignani \cite{C01} has analyzed the simple shear flow problem of a monocomponent granular gas considering the more refined version of the Maxwell model indicated in Sec.\ \ref{sec2}.  The results obtained by Cercignani \cite{C01} in the (steady) shear flow 
problem for a three dimensional system 
leads to the absence of the normal stress differences ($P_{xx}^*=P_{yy}^*=P_{zz}^*=1$) and the 
shear stress is given by 
\begin{equation}
\label{3.18}
P_{xy}^*=-\sqrt{3\frac{1-\alpha}{3-\alpha}}.
\end{equation}

The elements of the pressure tensor of IHS have been explicitly obtained from  the Boltzmann equation 
in the first Sonine approximation.\cite{MG02,G02}  For the sake of completeness, their expressions 
are displayed in Appendix \ref{appB} [cf. Eqs.\ (\ref{b5}--(\ref{b7})]. It is apparent that these expressions differ from those found here, Eqs.\ (\ref{3.14})--(\ref{3.16}). However, for practical purposes, the discrepancies between both interaction models are quite small, even for moderate values of $\alpha$. As an ilustration, Fig.  \ref{fig1} shows a comparison of the results obtained for IMM and IHS for the (reduced) elements of the pressure tensor for $d=3$ . 
The results (symbols) obtained from Monte Carlo simulations for IHS \cite{MG02} as well as the predictions for the shear stress given by Eq.\ (\ref{3.18}) are also included. We observe that the discrepancies between IHS and IMM are very small, although they slightly increase as the restitution coefficient decreases. 
In addition, the agreement  of both interaction models  with simulation data is quite good. On the other hand, the results derived in ref.\ \onlinecite{C01} exhibit significant discrepancies with Monte Carlo results, except in the low-dissipation region.

\subsection{Tracer limit}
\label{sec3.2}

When one of the species, say species 1, has a vanishing mole fraction, it acts as a tracer species.  In this case, the state of the excess component 2 is not affected by the presence of the tracer particles and so the expression of ${\sf P}_2^*$ is the same as the one obtained in the single gas case, Eqs.\ (\ref{3.14})--(\ref{3.16}), with $\alpha=\alpha_{22}$. The corresponding expressions for the elements of the tracer pressure tensor ${\sf P}_1^*$ can be 
obtained from Eq.\ (\ref{3.5}) by taking carefully the limit $x_1\to 0$. After some algebra, one gets
\begin{equation}
\label{3.19}
P_{1,yy}^*=P_{1,zz}^*=\cdots=P_{1,dd}^*=-\frac{F+HP_{2,yy}^*}{G},
\end{equation} 
\begin{equation}
\label{3.20}
P_{1,xy}^*=\frac{a^*P_{1,yy}^*-HP_{2,xy}^*}{G},
\end{equation} 
\begin{equation}
\label{3.21}
P_{1,xx}^*=d\gamma-(d-1)P_{1,yy}^*,
\end{equation} 
where 
\begin{equation}
\label{3.22}
F=\frac{2}{d+2}\mu_{12}^{3/2}\mu_{21}\left(\frac{1+\theta}{\theta}\right)^{3/2}(1+\alpha_{12})^2,
\end{equation}
\begin{equation}
\label{3.23}
G=\frac{4}{d(d+2)}\sqrt{\mu_{12}}\mu_{21}\left(\frac{1+\theta}{\theta}\right)^{1/2}(1+\alpha_{12})
\left[\mu_{21}(1+\alpha_{12})-d-2\right],
\end{equation}
\begin{equation}
\label{3.24}
H=\frac{4}{d(d+2)}\mu_{12}^{3/2}\mu_{21}\left(\frac{1+\theta}{\theta}\right)^{1/2}(1+\alpha_{12})^2,
\end{equation}
and the reduced shear rate is
\begin{equation}
\label{3.18bis}
{a^*}^2=\left(\frac{\sigma_2}{\sigma_{12}}\right)^{2(d-1)}\frac{2(d+1-\alpha_{22})^2(1-\alpha_{22}^2)}{d^2(d+2)}.
\end{equation}
Here, $\theta=m_1T_2/m_2T_1$ is the mean square velocity of the gas particles relative
to that of the tracer particles. The temperature ratio $\gamma$ is obtained from Eq.\ (\ref{3.12}) where the (reduced) cooling rates $\zeta_2^*$ and $\zeta_1^*$ in the tracer limit are given by  
\begin{equation}
\label{3.25}
\zeta_2^*=\frac{\sqrt{2\mu_{12}}}{d}\left(\frac{\sigma_{2}}{\sigma_{12}}\right)^{d-1}(1-\alpha_{22}^2),
\end{equation}
\begin{equation}
\label{3.26}
\zeta_1^*=\frac{4}{d}\mu_{21}\sqrt{\mu_{12}}\left(\frac{1+\theta}{\theta}\right)^{1/2}(1+\alpha_{12})
\left[1-\frac{\mu_{21}}{2}(1+\theta)(1+\alpha_{12})\right].
\end{equation}

\subsection{General case: comparison with IHS results }
\label{sec3.3}

Here, I compare the results obtained for IMM with those found for 
IHS by using the leading Sonine approximation \cite{MG02} as well as by performing Monte Carlo 
simulations.\cite{MG02,MG02bis} Specifically, I want to explore the dependence of the temperature ratio $\gamma$ 
and the nonzero elements of ${\sf P}^*$ on the restitution coefficients $\alpha_{rs}$, the mass ratio $m_1/m_2$, the 
concentration ratio $n_1/n_2$, and the ratio of sizes $\sigma_1/\sigma_2$. Given the high dimensionality of the 
parameter space, for the sake of concreteness, henceforth I will assume that the disks or spheres are made of 
the same material, i.e., $\alpha_{11}=\alpha_{22}=\alpha_{12}\equiv \alpha$.  This reduces the parameter space to four quantities.

The temperature ratio $\gamma=T_1/T_2$ measures the breakdown of energy equipartition. Due to the coupling between dissipation and the shear rate in the (steady) simple shear flow state [cf. Eq.\ (\ref{3.10})], the mixture departures from 
the Navier-Stokes regime as the restitution coefficient $\alpha$ decreases. Thus, energy nonequipartition is expected except when $\alpha\to 1$. The coexistence of two granular temperatures has been recently confirmed by molecular dynamics simulations carried out in sheared granular 
mixtures of hard disks\cite{CH02} and by some experiments \cite{WP02} of vibrated granular mixtures. Related findings have been reported by 
some authors by using kinetic theory tools in the freely cooling state for IHS \cite{GD99bis,DG01} and IMM. \cite{MP02,NK02}    

As expected, the results obtained here show that in general the kinetic temperatures of the mixture are different 
($\gamma\neq 1$). The dependence of $\gamma$ on the parameters of the problem is very complex. In Fig.\ \ref{fig2} the temperature ratio $T_1/T_2$ is plotted versus the restitution coefficient $\alpha$ for an equimolar mixture ($n_1=n_2$) of hard spheres ($d=3$) 
with $\sigma_1/\sigma_2=1$ and two different values of the mass ratio: $m_1/m_2=2$ and 10. We clearly see that the predictions for IMM agree quite well with the ones obtained for IHS in the first Sonine approximation and that the agreement of both approaches with simulation is excellent, even for moderate dissipation.  It is also apparent that the extent of the equipartition violation is greater when the mass disparity 
is large.  To illustrate this point, in Fig.\ \ref{fig3}  we consider an equimolar  mixture of hard disks ($d=2$) of constant density, 
i.e., $m_1/m_2=(\sigma_1/\sigma_2)^2$.  Three different values of the restitution coefficient have been studied:  $\alpha=0.95, 0.9,$ and 0.8. Again, the IMM results reproduce quite well the IHS predictions (theory and simulation) over the entire range of 
values of size and mass ratios considered. Further, the behavior of $T_1/T_2$ obtained here for dilute systems  is qualitatively similar to that found in molecular dynamics simulations\cite{CH02} for finite-density systems. Thus, for instance, at a given value of $\alpha$, the granular energy of the larger particle (say for instance, species 1) increases relative to that of the smaller particle as the ratio $\sigma_1/\sigma_2$ (or equivalently, $m_1/m_2$) increases.

The dependence of the nonzero elements of the pressure tensor ${\sf P}^*$ on the restitution coefficient $\alpha$ is illustrated in Figs.\ \ref{fig4} and \ref{fig5} for two different mixtures. In Fig.\ \ref{fig4}, we consider a mixture of hard spheres ($d=3$) with $n_1/n_2=\sigma_1/\sigma_2=1$ and $m_1/m_2=2$, while Fig.\ \ref{fig5} corresponds  to a mixture of hard disks ($d=2$) with $n_1/n_2=1$, $m_1/m_2=(\sigma_1/\sigma_2)^2=10$. We see that the agreement between the results derived for IMM and IHS is again in general quite good, especially in the case of the shear stress $P_{xy}^*$ which is the most relevant rheological property in a shear flow problem. In fact, in the two-dimensional case, the exact results obtained here for IMM are closer to the simulation results  than those obtained by using the first Sonine approximation for IHS.\cite{MG02} It must be remarked that  both theories only predict normal stress differences in the plane of shear flow ($P_{xx}^*>P_{yy}^*=P_{zz}^*=\cdots =P_{dd}^*$), while the simulation results also shows anisotropy in the plane orthogonal to the flow velocity 
($P_{zz}^*>P_{yy}^*$). However, these relative normal stress differences are very small and tend to zero as the restitution coefficient decreases.

\section{Diffusion of impurities under granular shear flow}
\label{sec4}

The second problem analyzed in this paper refers to the diffusion of impurities  in a granular Maxwell gas subjected to the simple shear flow.  This problem has also been recently analyzed by the author \cite{G02} in the framework of IHS. 
In the tracer limit ($x_1\to 0$), the state of the excess component is not  affected 
by the presence of the tracer particles so that  its velocity distribution function $f_2$ verifies a closed Boltzmann 
equation. As a consequence,  the pressure tensor ${\sf P}_2^*$ of the granular gas is given by Eqs.\  (\ref{3.14})--(\ref{3.16}).  Further, given that the mole
fraction of the tracer particles $x_1$ is very small, one can neglect their mutual interactions in the kinetic equation for $f_1$ and assume that $f_1$ verifies a (linear) Boltzmann-Lorentz equation.  
The diffusion process is induced in the system by the presence of a {\em weak} concentration gradient $\nabla x_1$. Given that the strength of the shear rate $a$ is arbitrary, the mass flux  ${\bf j}_1$ (which is generated by the gradient $\nabla x_1$) can be modified by the presence of the shear flow. Under these conditions, a diffusion tensor ${\sf D}$ is required to describe the diffusion process instead of a single diffusion coefficient.  The determination of the nonzero elements of this tensor is the goal of this Section.

In the local Lagrangian frame, the kinetic equation governing the evolution 
of the velocity distribution function $f_1$ reads
\begin{equation}
\label{4.1}
\frac{\partial}{\partial t}f_1-a_{ij} V_{j}\frac{\partial}{\partial 
V_{i}}f_1+(V_i+a_{ij}r_{j})\frac{\partial}{\partial 
r_i}f_1=J_{12}[{\bf V}|f_1,f_2]\;,
\end{equation}
where here the derivative $\partial/\partial r_i$ is taken at constant ${\bf V}$. Assuming that the mole fraction $x_1$ is slightly {\em nonuniform}, we solve Eq.\ (\ref{4.1}) by means of a 
perturbation expansion around a nonequilibrium state with arbitrary shear rate. Thus, we write
\begin{equation}
\label{4.3}
f_1=f_1^{(0)}+f_1^{(1)}+\ldots, 
\end{equation}
where $f_1^{(k)}$ is of order $k$ in $\nabla x_1$ but applies for {\em arbitrary} degree of dissipation since this distribution retains all the orders in $a^*$. The solution (\ref{4.3}) qualifies as a normal solution since all the space and time dependence of $f_1$ occurs entirely
through $x_1({\bf r};t)$ and their gradients. The zeroth-order approximation $f_1^{(0)}$ corresponds to the
simple shear flow distribution but taking into account now the local 
dependence on the mole fraction $x_1$. Although the explicit form of 
$f_1^{(0)}$ is not exactly known, only the knowledge of its second-degree 
moments is necessary to get the diffusion tensor. The explicit expressions of the (reduced) pressure 
tensor ${\sf P}_1^*$ are given by Eqs.\ (\ref{3.19})--(\ref{3.24}).

The kinetic equation for $f_1^{(1)}$ can be obtained from the Boltzmann-Lorentz
equation (\ref{4.1}) by collecting all the terms of first order in $\nabla x_1$: 
\begin{equation}
\label{4.4}
\frac{\partial}{\partial t}f_1^{(0)}-a_{ij} V_{j}\frac{\partial}{\partial 
V_{i}}f_1^{(1)}+(V_i+a_{ij}r_j)\frac{\partial}{\partial 
r_i}f_1^{(0)}=J_{12}[{\bf V}|f_1^{(1)},f_2]\;.
\end{equation} 
The determination of the diffusion tensor from Eq.\ (\ref{4.4}) follows similar mathematical steps as those made in the case of IHS. I refer the reader to ref.\ \onlinecite{G02} for more specific details and here only some intermediate results will be displayed. Using the balance equation for the mass density of impurities, Eq.\ (\ref{4.4}) becomes  
\begin{equation}
\label{4.7}
\left(a_{ij} V_{j}\frac{\partial}{\partial V_{i}}+\Lambda\right)f_1^{(1)}=
\frac{\partial f_1^{(0)}}{\partial x_1}\left({\bf V}\cdot \nabla 
x_1\right)\;,
\end{equation}
where $\Lambda$ is the Boltzmann-Lorentz collision operator 
\begin{equation}
\label{4.8}
\Lambda f_1^{(1)}=J_{12}[{\bf V}|f_1^{(1)},f_2].
\end{equation}
The mass flux  ${\bf j}_1^{(1)}$ is defined as 
\begin{equation} 
{\bf j}_{1}^{(1)}=m_{1}\int d{\bf v}\,{\bf V}\,f_{1}^{(1)}.
\label{4.8bis} 
\end{equation} 
The components of ${\bf j}_1^{(1)}$ can be obtained by multiplying Eq.\ (\ref{4.7}) by $m_1{\bf V}$ 
and integrating over ${\bf V}$. The result can be written as  
\begin{equation}
\label{4.9}
\left({\sf a}+\lambda \openone \right)\cdot {\bf j}_1^{(1)}=-p_2 {\sf P}_1^*\cdot \nabla x_1,
\end{equation}
where use has been made of Eq.\ (\ref{2.15}) (with ${\bf j}_2={\bf 0}$) and 
$\openone$ is the $d\times d$ unit tensor. The coefficient $\lambda$ is
\begin{eqnarray}
\label{4.10}
\lambda&=&\frac{w_{12}}{d}\mu_{21}(1+\alpha_{12})\nonumber\\
&=& \frac{2}{d}\frac{\pi^{(d-1)/2}}{\Gamma(d/2)}
n_2\mu_{21}\sigma_{12}^{d-1}(2T_2/m_2)^{1/2}
(1+\alpha_{12})\left(\frac{1+\theta}{\theta}\right)^{1/2}.
\end{eqnarray}
Note that, according to Eq.\ (\ref{2.17}),  $w_{12}$ is still given by  Eq.\ (\ref{3.13bis}) since 
${\bf j}_2={\bf 0}$. The solution to Eq.\ (\ref{4.9}) can be written in the form 
\begin{equation}
\label{4.11}
{\bf j}_1^{(1)}=-{\sf D}\cdot \nabla x_1,
\end{equation} 
with the elements of the diffusion tensor ${\sf D}$ given by 
\begin{equation}
\label{4.12}
D_{ij}=\frac{p_2}{\lambda}\left(\delta_{ik}-\frac{a_{ik}}{\lambda}\right) P_{1,kj}^{*}.
\end{equation}
Equation (\ref{4.12}) provides an explicit expression  of the tracer diffusion tensor of a granular Maxwell mixture under shear flow. In the absence of shear rate ($a=0$), as well as in the elastic limit ( $\alpha_{22}=\alpha_{12}=1$), one has $D_{ij}=D_0\delta_{ij}$, where 
\begin{equation}
\label{4.13}
D_0=\frac{d}{4\sqrt{2}}\frac{\Gamma(d/2)}{\pi^{(d-1)/2}
\sigma_{12}^{d-1}}\sqrt{\frac{m_1(m_1+m_2)}{m_2}T_2}
\end{equation}
is the tracer diffusion coefficient of a gas of Maxwell molecules.\cite{CC70}  As the 
restitution coefficient decreases, rheological effects become important and 
the elements of the diffusion tensor are different from the one obtained in 
the equilibrium case. The dependence of the diffusion coefficients 
on the restitution coefficients $\alpha_{22}$ and $\alpha_{12}$ as well as 
on the mass ratio $m_1/m_2$ and the size ratio $\sigma_1/\sigma_2$ is highly nonlinear. 
As happens for elastic fluids,\cite{G} Eq.\ (\ref{4.12}) shows that diffusion 
under simple shear flow is a very complex problem due basically to the 
anisotropy induced in the system by the shear flow.

To illustrate the dependence of the elements of the diffusion tensor on the parameters 
of the problem, let us consider the three dimensional case ($d=3$). 
According to Eq.\ (\ref{4.12}), 
$D_{xz}=D_{zx}=D_{yz}=D_{zy}=0$, in agreement with the symmetry of the 
problem. Consequently, there are five relevant elements: the three diagonal 
and two ($D_{xy}$, $D_{yx}$) off-diagonal elements. In general,
$D_{xx} \neq D_{yy}=D_{zz}$ and $D_{xy} \neq D_{yx}$. The off-diagonal elements are negative and measure cross effects in the diffusion of impurities in a sheared granular gas. 
The results obtained \cite{G02} for IHS  [cf.  Eqs.\ (\ref{b9})--(\ref{b16}) of Appendix \ref{appB}] also predict anisotropy in the plane perpendicular 
to the shear flow ($D_{yy} \neq D_{zz}$), although this anisotropy is much smaller than the one observed in the plane of shear flow (as measured by the difference $D_{xx}-D_{yy}$).   
In Fig.\ \ref{fig6}, the self-diffusion problem is considered, i. e., when the tracer particles are mechanically equivalent to the gas particles. This figure shows $D_{xx}^*-D_{yy}^*$,  
$(D_{xx}^*+D_{yy}^*+D_{zz}^*)/3\equiv (1/3)D_{kk}^*$, $D_{xy}^*$, and 
$D_{yx}^*$ as functions of the restitution coefficient $\alpha\equiv 
\alpha_{12}=\alpha_{22}$. Here, $D_{ij}^*\equiv D_{ij}/D_0$, with $D_0$ 
given by Eq.\ (\ref{4.13}). We see that the deviation from the functional 
form for elastic collisions is quite important even for moderate 
dissipation. The figure also shows 
that the anisotropy of the system, as measured by the difference 
$D_{xx}^*-D_{yy}^*$ grows with the inelasticity.  It is worthwhile remarking that the general qualitative behavior of the self-diffusion tensor on dissipation agrees well with molecular dynamics simulations carried out by Campbell, \cite{C97} at least for the lowest solid fraction considered in his simulations. With respect to the comparison with the results for IHS, \cite{G02} we see again  good agreement although the discrepancies between both interaction models slightly
increases as $\alpha$ decreases. In fact, these discrepancies are more significant than those found for the rheological properties in the simple shear flow problem.  Finally, Fig.\ \ref{fig7} shows the dependence of $D_{ij}^*$ on $\alpha$  in the case $m_1/m_2=2$ and $\sigma_1/\sigma_2=2$ for  IMM and IHS. Similar conclusions to those made in the self-diffusion case are obtained here for 
mixtures of species with different (but not disparate) mass and size. It must be pointed out that the  discrepancies for the diffusion tensor between both interaction  models turn out to be significant as the disparity of masses or sizes increases.

\section{Concluding remarks}
\label{sec5}

Needless to say, the analysis of nonlinear transport phenomena (i.e., when the hydrodynamic gradients are not necessarilly small ) from the Boltzmann equation for inelastic hard spheres (IHS) is a formidable task. For this reason, to get explicit results one usually considers the leading order in a Sonine  polynomial expansion of the velocity distribution function. All the above difficulties increase when one studies polydisperse systems (granular mixtures) since not only is the number of transport coefficients larger than for a monomponent gas but they are also functions of more parameters such as the concentrations, masses, sizes, and the coefficients of restitution. A possible way to partially overcome these problems is to consider  
interaction models (such as the inelastic Maxwell model) that  simplify the mathematical structure of the Boltzmann collision operator. For elastic fluids, this strategy has been shown to be very fruitful since the exact results obtained for Maxwell molecules for many interesting transport properties (both linear and nonlinear)  agrees quite well with hard spheres and other interaction potentials when they are convenienlty nondimensionalized. \cite{SG95}

In the Boltzmann equation for inelastic Maxwell models (IMM), the collision rate of IHS is replaced by an effective collision rate independent of the two colliding particles.  The simplicity of this  interaction model allows us for instance, to get exactly the moments of the Boltzmann collision operator without the explicit knowledge of the velocity distribution function. The cost of sacrificing physical realism  can be compensated by the amount of exact analytical results obtained from IMM. However, in order to assess the degree of usefulness of IMM as a prototype model for the description of granular media, one needs to compare the predictions based on IMM with those obtained in the framework of the IHS model. In this paper I have carried out such a comparison at the level of nonlinear transport in two different but related nonequilibrium problems. First,  I have explicitly determined  the rheological properties (pressure tensor) of a binary inelastic Maxwell mixture under simple shear flow. Then, by taking the above solution as the reference state, I have obtained the diffusion tensor of impurities immersed in a sheared inelastic Maxwell gas. In the context of IMM, both expressions are exact in the sense that they apply for arbitrary values of the restitution coefficients and the parameters characterizing the mixture (masses, sizes, and concentrations).

Previous studies of such nonlinear transport coefficients  have been carried out  \cite{MG02,G02} by the author in the case of IHS. As said above, these results \cite{MG02,G02} are approximate since they are obtained by retaining the first terms in a Sonine polynomial expansion of the velocity distribution functions.  However, the accuracy of these results has been confirmed by Monte Carlo simulations \cite{MG02,MG02bis} of the Boltzmann equation. To compare the results derived from IMM and IHS,  the collision frequencies $w_{rs}$ appearing in the Boltzmann equation for IMM [see Eq.\ (\ref{2.2})] need to be fixed. Here, to make contact with IHS, I have chosen $w_{rs}$ to reproduce the cooling rates $\zeta_{rs}$ of IHS. With this choice, the results derived here for IMM for the pressure tensor and the diffusion tensor compare quite well with the ones previously derived for IHS over a wide range of values of dissipation and parameters of the mixture.  
Given the high dimensionality of the parameter space explored,  the comparison carried out in this paper can be considered as a good test to gauge the reliability of IMM to reproduce the main trends observed for IHS in the context of granular mixtures.

Finally, it must be noted that the good agreement found here contrasts with the results recently derived for the Navier-Stokes (NS) transport coefficients,\cite{S02}  where it was shown that the dependence of the transport coefficients of IMM on the restitution coefficient only compares well at a qualitatively level with the one obtained for IHS. However, the situation analyzed in ref.\ \onlinecite{S02} is completely different to the one studied here since the Navier-Stokes coefficients were obtained in the first order of the Chapman-Enskog expansion around a time-dependent state (homogeneous cooling state). In this sense, it seems that 
the results found in this paper for IMM may be sufficiently representative of the trends observed for IHS in some particular situations (such as the simple shear flow problem), and this agreement  must be taken with caution when one explores other situations.  This fact stimulates the search for exact solutions for IMM which can be confronted with the results obtained for IHS by using approximate analytical methods and computer simulations.  

\acknowledgments

Partial support from the Ministerio de Ciencia y Tecnolog\'{\i}a (Spain) through Grant No. BFM2001-0718 is acknowledged.

\appendix

\section{Collisional moments in the inelastic Maxwell model}
\label{appA}

In this Appendix I will derive the collisional moments appearing in the main text. Let us 
consider the general collisional integral 
\begin{equation}
\label{a1}
I_{rs}[F]=\int d{\bf v}F({\bf V}) J_{rs}\left[ f_{r},f_{s}\right] 
\end{equation}
A useful identity for an arbitrary function $F({\bf V})$ is given by  
\begin{equation} 
\label{a2}
I_{rs}[F]=\frac{w _{rs}}{n_s\Omega_d}
\int \,d{\bf v}_{1}\,\int \,d{\bf v}_{2}f_{r}({\bf V}_{1})f_{s}({\bf V}_{2})  
\int d\widehat{\bbox {\sigma}}\,\left[F({\bf V}_1'')-F({\bf V}_1)\right],
\end{equation} 
with  
\begin{equation}
\label{a3} 
{\bf V}_{1}^{^{\prime \prime}}={\bf V}_{1}-\mu _{sr}(1+\alpha _{rs})(  
\widehat{\bbox {\sigma }}\cdot {\bf g}_{12})\widehat{\bbox {\sigma}},
\end{equation} 
and ${\bf g}_{12}={\bf V}_1-{\bf V}_2$. Now we particularize to $F({\bf V})=m_r{\bf V}$.  Using (\ref{a2}), one gets
\begin{equation}
\label{a4}
I_{rs}[m_r{\bf V}]=-\frac{w _{rs}}{n_s\Omega_d}m_r\mu_{sr}(1+\alpha _{rs})
\int \,d{\bf v}_{1}\,\int \,d{\bf v}_{2}f_{r}({\bf V}_{1})f_{s}({\bf V}_{2})  
\int d\widehat{\bbox {\sigma}}(\widehat{\bbox {\sigma }}\cdot {\bf g}_{12})
\widehat{\bbox {\sigma}}\;. 
\end{equation}
To perform the angular integration, one needs the result \cite{EB02}
\begin{equation}
\label{a5}
\int d\widehat{\bbox {\sigma}}\,(\widehat{\bbox {\sigma}}\cdot {\bf g}_{12})^k
\widehat{\bbox {\sigma}}=\beta_{k+1} g_{12}^{k-1}{\bf g}_{12},
\end{equation}
where 
\begin{equation}
\label{a7}
\beta_k=\Omega_d\pi^{-1/2}\frac{\Gamma\left(\frac{d}{2}\right)
\Gamma\left(\frac{k+1}{2}\right)}{\Gamma\left(\frac{k+d}{2}\right)}.
\end{equation}
Thus, the integration over $\widehat{\bbox {\sigma}}$ in Eq.\ (\ref{a4}) 
leads to 
\begin{eqnarray}
\label{a8}
I_{rs}[m_r{\bf V}]&=&-\frac{w _{rs}}{n_sd}m_r\mu_{sr}(1+\alpha _{rs})
\int \,d{\bf v}_{1}\,\int \,d{\bf v}_{2}f_{r}({\bf V}_{1})f_{s}({\bf V}_{2})  
\left({\bf V}_1-{\bf V}_2\right) \nonumber\\
&=& -\frac{w _{rs}}{\rho_sd}\mu_{sr}(1+\alpha _{rs})
\left(\rho_s{\bf j}_r-\rho_r{\bf j}_s\right),
\end{eqnarray}
where $\rho_r=m_r n_r$ and 
 \begin{equation}
\label{a9}
{\bf j}_r=\int d{\bf v} m_r {\bf V} f_r({\bf v}).
\end{equation}

Next, we particularize to $F({\bf V})=m_r{\bf V}{\bf V}$.  From the collision rule (\ref{a3}) it follows that
\begin{equation}
\label{a11}
{\bf V}_1''{\bf V}_1''-{\bf V}_1{\bf V}_1=
\mu_{sr}(1+\alpha_{rs})(\widehat{\bbox {\sigma }}\cdot {\bf g}_{12})
\left[\mu_{sr}(1+\alpha_{rs})(\widehat{\bbox {\sigma }}\cdot {\bf g}_{12})
\widehat{\bbox {\sigma }}\widehat{\bbox {\sigma }}-({\bf V}_1\widehat{\bbox {\sigma }}+
\widehat{\bbox {\sigma }}{\bf V}_1)\right].
\end{equation}
To perform the angular integration in (\ref{a8}), we need now the result
\begin{equation}
\label{a12}
\int d\widehat{\bbox {\sigma}}\,(\widehat{\bbox {\sigma}}\cdot {\bf g}_{12})^k \widehat{\bbox {\sigma}}\widehat{\bbox {\sigma}}
=\frac{\beta_{k}}{k+d} g_{12}^{k-2}\left(k{\bf g}_{12}{\bf g}_{12}+g_{12}^2
\openone\right),
\end{equation}
where $\openone$ is the $d\times d$ unit tensor. By using the identity  (\ref{a2}) and 
Eqs.\ (\ref{a11}) and (\ref{a12}), the collisional moment $I_{rs}[m_r{\bf V}{\bf V}]$ can be finally written as 
\begin{eqnarray}
\label{a13}
I_{rs}[m_r{\bf V}{\bf V}] &=&\frac{w _{rs}}{n_sd}m_r\mu_{sr}(1+\alpha _{rs})
\int \,d{\bf V}_{1}\,\int \,d{\bf V}_{2}f_{r}({\bf V}_{1})f_{s}({\bf V}_{2})  
\left[\frac{\mu_{sr}(1+\alpha _{rs})}{d+2}\left(2{\bf g}_{12}{\bf g}_{12}+g_{12}^2\openone\right)
\right. \nonumber\\
& & \left. -\left({\bf V}_1{\bf g}_{12}+{\bf g}_{12}{\bf V}_1\right)\right]
\nonumber\\
&=& -\frac{w _{rs}}{\rho_sd}\mu_{sr}(1+\alpha _{rs})\left\{2\rho_s{\sf P}_r-\left(
{\bf j}_r{\bf j}_s+{\bf j}_s{\bf j}_r\right)\right. \nonumber\\
& &-\frac{2}{d+2}\mu_{sr}(1+\alpha _{rs})\left[\rho_s{\sf P}_r+\rho_r{\sf P}_s-
\left({\bf j}_r{\bf j}_s+{\bf j}_s{\bf j}_r\right)\right.\nonumber\\
& & 
\left.\left.+\left[\frac{d}{2}\left(\rho_rp_s+\rho_sp_r\right)-{\bf j}_r\cdot {\bf j}_s\right]\openone
\right]\right\}.
\end{eqnarray}
Here, $p_r=\text{Tr}{\sf P}_r/d=n_rT_r$ with 
\begin{equation}
\label{a14}
{\sf P}_r=\int d{\bf v} m_r {\bf V} {\bf V}f_r({\bf v}).
\end{equation}

In the simple shear flow state, ${\bf j}_r={\bf 0}$, and the expression (\ref{a13}) for the collisional moment  $I_{rs}[m_r{\bf V}{\bf V}]$ becomes
\begin{equation}
\label{a15}
I_{rs}[m_r{\bf V}{\bf V}] =\frac{2}{d+2}\frac{w _{rs}}{\rho_sd}\mu_{sr}^2(1+\alpha _{rs})^2
\left[\left(1-\frac{d+2}{\mu_{sr}(1+\alpha_{rs})}\right)\rho_s{\sf P}_r+ \rho_r{\sf P}_s+
\frac{d}{2}\left(\rho_rp_s+\rho_sp_r\right)\openone\right].
\end{equation}
From Eq.\ (\ref{a15}) one can easily get the coefficients $A_{rs}$ and $B_{rs}$ appearing in 
Eqs.\ (\ref{3.6})--(\ref{3.9}). In the case of mechanically 
equivalent particles  ($m_1=m_2=m, \sigma_1=\sigma_2=\sigma, \alpha_{11}=\alpha_{22}=
\alpha_{12}=\alpha$), one has $T_1=T_2=T$, ${\sf P}_1={\sf P}_2={\sf P}$, and so the collisional moment $I_{rs}[m_r{\bf V}{\bf V}]\equiv I[m{\bf V}{\bf V}]$ reduces to
\begin{equation}
\label{a18}
I[m{\bf V}{\bf V}] =\frac{1}{2(d+2)}w (1+\alpha)\left[(1+\alpha)p\openone-\frac{2}{d}(d+1-\alpha){\sf P}\right].
\end{equation}
This expression coincides with the one previously derived in the monocomponent case.\cite{S02}

\section{Some expressions for inelastic hard spheres}
\label{appB}

The shear flow problem for a binary mixture of IHS was studied in refs. \ \onlinecite{MG02} and \onlinecite{G02}. I give here some expressions obtained for this interaction model in the first Sonine approximation that have been used along the paper to compare them with the results derived for IMM. 

For IHS the expressions of the coefficients $A_{11}$, $A_{21}$, $B_{11}$, and $B_{12}$ 
appearing in Eqs.\  (\ref{3.5}) are given by  
\begin{equation}
\label{b1}
A_{11}=\frac{\sqrt{2}}{d(d+2)}x_1\left(\frac{\sigma_1}{\sigma_{12}}\right)^{d-1}
\theta_1^{-1/2}(1+\alpha_{11})\left[\alpha_{11}(d-1)+d+1\right]\nu,
\end{equation}
\begin{equation}
\label{b2}
A_{12}=\frac{2}{d(d+2)}x_1\mu_{12}\theta_2^{-1/2}\left(\frac{\theta_1+\theta_2}{\theta_1}\right)^{3/2}
(1+\alpha_{12})\left[\frac{2\theta_2}{\theta_1+\theta_2}+\mu_{21}(d-1)(1+\alpha_{12})\right]\nu,
\end{equation}
\begin{eqnarray}
\label{b3}
B_{11}&=&\frac{\sqrt{2}}{d(d+2)}x_1\left(\frac{\sigma_1}{\sigma_{12}}\right)^{d-1}\mu_{21}\theta_1^{1/2}
(1+\alpha_{11})(2d+3-3\alpha_{11})\nu \nonumber\\
& & +\frac{2}{d(d+2)}x_2\mu_{21}^2\left(\frac{\theta_1}{\theta_2(\theta_1+\theta_2)}\right)^{1/2}
\left\{\left[2d-3\mu_{21}(1+\alpha_{12})\right] (\theta_1+\theta_2)+2(2\theta_1+3\theta_2)\right\}\nu,
\end{eqnarray}
\begin{equation}
\label{b4}
B_{12}=-\frac{2}{d(d+2)}x_1\mu_{12}^2\left(\frac{\theta_2^3}{\theta_1(\theta_1+\theta_2)}\right)^{1/2}
\left[3\mu_{21}\frac{\theta_1+\theta_2}{\theta_2}(1+\alpha_{12})-2\right]\nu.
\end{equation}
Here, $\theta_1$ and $\theta_2$ are defined by Eq.\ (\ref{2.17bis}) while $\nu$ is the effective collision frequency defined by Eq.\ (\ref{3.9.1}). From Eqs.\ (\ref{b1})--(\ref{b4}) and
their counterpart for species 2, the temperature ratio and the elements of the pressure tensor 
can be easily obtained. In the special case of mechanically equivalent particles, Eqs.\ (\ref{b1})--(\ref{b4}) lead to the following expressions for the nonzero elements of the reduced presure tensor ${\sf P}^*$: 
\begin{equation}
\label{b5}
P_{yy}^*=P_{zz}^*=\cdots=P_{dd}^*=\frac{d+1+(d-1)\alpha}{2d+3-3\alpha},
\end{equation}
\begin{equation}
\label{b6}
P_{xy}^*=-\sqrt{\frac{d(d+2)[d+1+(d-1)\alpha](1-\alpha)}{2(2d+3-3\alpha)^2}},
\end{equation}
\begin{equation}
\label{b7}
P_{xx}^*=d-(d-1)P_{yy}^*.
\end{equation}

Tracer diffusion under shear flow was analyzed in ref.\ \onlinecite{G02} for IHS. The expression of the diffusion tensor is
\begin{equation}
\label{b9}
{\sf D}=p_2\left({\sf a}+{\bbox {\Omega}}\right)^{-1}\cdot {\sf P}_{1}^{*},
\end{equation}
where the tensor ${\bbox {\Omega}}$ is 
\begin{equation}
\label{b9bis} 
{\bbox {\Omega}}=\lambda 
\left[\openone+\frac{1}{d+2}\frac{\theta}{1+\theta} \left(
{\sf P}_{2}^*-\openone\right)\right].
\end{equation}
Here, $\lambda$ is given by Eq.\ (\ref{4.10}) and the pressure tensor ${\sf P}_2^*$ of the excess component is given by Eqs.\ (\ref{b5})--(\ref{b7}). The nonzero elements  of ${\sf P}_1^{*}$ can be written in the form given by Eqs.\ (\ref{3.19})--(\ref{3.21}) but the functions $F$, $G$, and $H$ are  
\begin{equation}
\label{b13}
F=\frac{4}{d(d+2)}\mu_{12}^{3/2}\left(\frac{1+\theta}{\theta^3}\right)^{1/2}(1+\alpha_{12})
\left[1+\frac{\mu_{21}}{2}(d-1)(1+\theta)(1+\alpha_{12})\right],
\end{equation}
\begin{eqnarray}
\label{b14}
G&=&-\frac{2}{d(d+2)}\sqrt{\mu_{12}}\mu_{21}\left(\frac{1}{\theta(1+\theta)}\right)^{1/2}
(1+\alpha_{12})\nonumber\\
& & \times
\left\{2[(d+2)\theta+d+3]-3\mu_{21}
(1+\theta)(1+\alpha_{12})\right\},
\end{eqnarray}
\begin{equation}
\label{b15}
H=\frac{2}{d(d+2)}\mu_{12}^{3/2}\left(\frac{1}{\theta(1+\theta)}\right)^{1/2}
(1+\alpha_{12})\left[3\mu_{21}(1+\theta)(1+\alpha_{12})-2\right],
\end{equation}
and the reduced shear rate is 
\begin{equation}
\label{b16}
{a^*}^2=\left(\frac{\sigma_2}{\sigma_{12}}\right)^{2(d-1)}
\frac{\mu_{12}}{d(d+2)} \frac{(1+\alpha_{22})(2d+3-3\alpha_{22})^2(1-\alpha_{22}^2)}{d+1+(d-1)\alpha_{22}}.
\end{equation}

\begin{figure}
\caption{Plot of the reduced elements of the pressure tensor ${\sf P}^*$ as functions of the restitution coefficient for a three-dimensional single gas.  The solid lines are the results derived here for IMM, the dashed lines correspond  to the results obtained for IHS from the Sonine approximation, and the symbols refer to Monte Carlo simulations for IHS. \protect{\cite{MG02}} The dotted line is the result given by Eq.\ (\ref{3.18}).}
\label{fig1}
\end{figure}

\begin{figure}
\caption{Plot of the temperature ratio $T_1/T_2$ versus the restitution coefficient in the three-dimensional case for $n_1/n_2=1$, $\sigma_1/\sigma_2=1$, and two different values of the mass ratio $m_1/m_2$:  $m_1/m_2=2$ and $m_1/m_2=10$ .  The solid lines are the results derived for IMM, the dashed lines correspond  to the results obtained for IHS from the Sonine approximation, and the symbols refer to Monte Carlo simulations for IHS. \protect{\cite{MG02}}}
\label{fig2}
\end{figure}

\begin{figure}
\caption{Plot of the temperature ratio $T_1/T_2$ as a function of the size ratio $\sigma_1/\sigma_2=(m_1/m_2)^{1/2}$ in the two-dimensional case for $n_1/n_2=1$ and three  different values of the restitution coefficient $\alpha$:  $\alpha=0.95$,  $\alpha=0.9$, and  $\alpha=0.8$.  The solid lines are the results derived for IMM, the dashed lines correspond  to the results obtained for IHS from the Sonine approximation, and the symbols refer to Monte Carlo simulations for IHS.\protect{\cite{MG02bis}}}
\label{fig3}
\end{figure}

\begin{figure}
\caption{Plot of the reduced elements of the pressure tensor ${\sf P}^*$ as functions of the restitution coefficient in the three-dimensional case for $n_1/n_2=1$, $\sigma_1/\sigma_2=1$,  and $m_1/m_2=2$. The solid lines are the results derived for IMM, the dashed lines correspond  to the results obtained for IHS from the Sonine approximation, and the symbols refer to Monte Carlo simulations for IHS.\protect{\cite{MG02}}}
\label{fig4}
\end{figure}

\begin{figure}
\caption{Plot of the reduced elements of the pressure tensor ${\sf P}^*$ as functions of the 
restitution coefficient $\alpha$ in the two-dimensional case for $n_1/n_2=1$,  and $\sigma_1/\sigma_2=(m_1/m_2)^{1/2}=10$.  The solid lines are the results derived for IMM, the dashed lines correspond  to the results obtained for IHS from the Sonine approximation, and the symbols refer to Monte Carlo simulations for IHS.\protect{\cite{MG02}}}
\label{fig5}
\end{figure}

\begin{figure}
\caption{Dependence of the diagonal and off-diagonal elements of the reduced self-diffusion tensor ${\sf D}^*$ on the restitution coefficient $\alpha$ in the three-dimensional case. The solid lines are the results derived for IMM while the dashed lines correspond  to the results obtained for IHS from the Sonine approximation. }
\label{fig6}
\end{figure}

\begin{figure}
\caption{Dependence of the diagonal and off-diagonal elements of the reduced diffusion tensor ${\sf D}^*$ on the restitution coefficient $\alpha$ in the three-dimensional case for $\sigma_1/\sigma_2=1$ and  $m_1/m_2=2$.  The solid lines are the results derived for IMM while the dashed lines correspond  to the results obtained for IHS from the Sonine approximation. }
\label{fig7}
\end{figure}

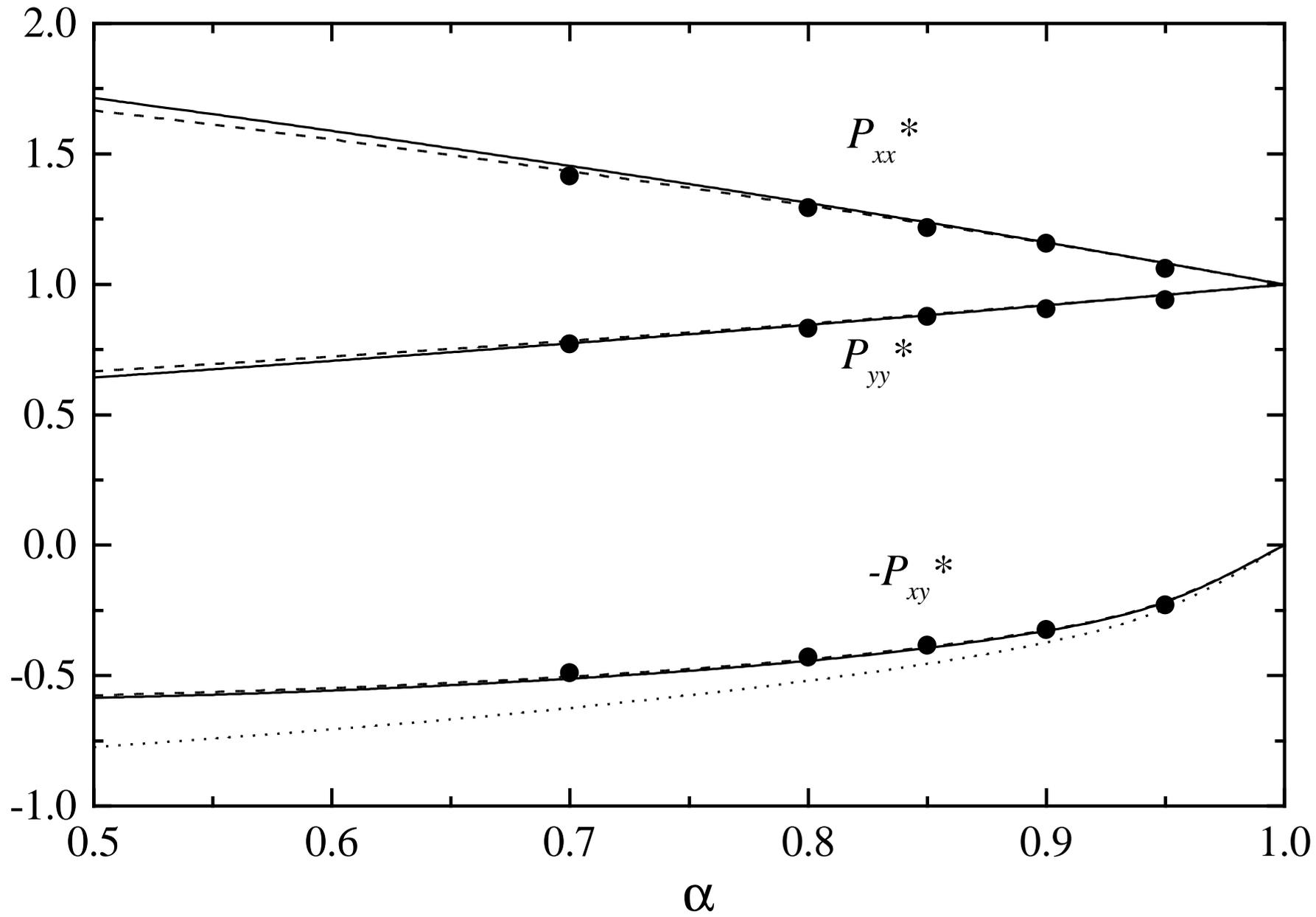
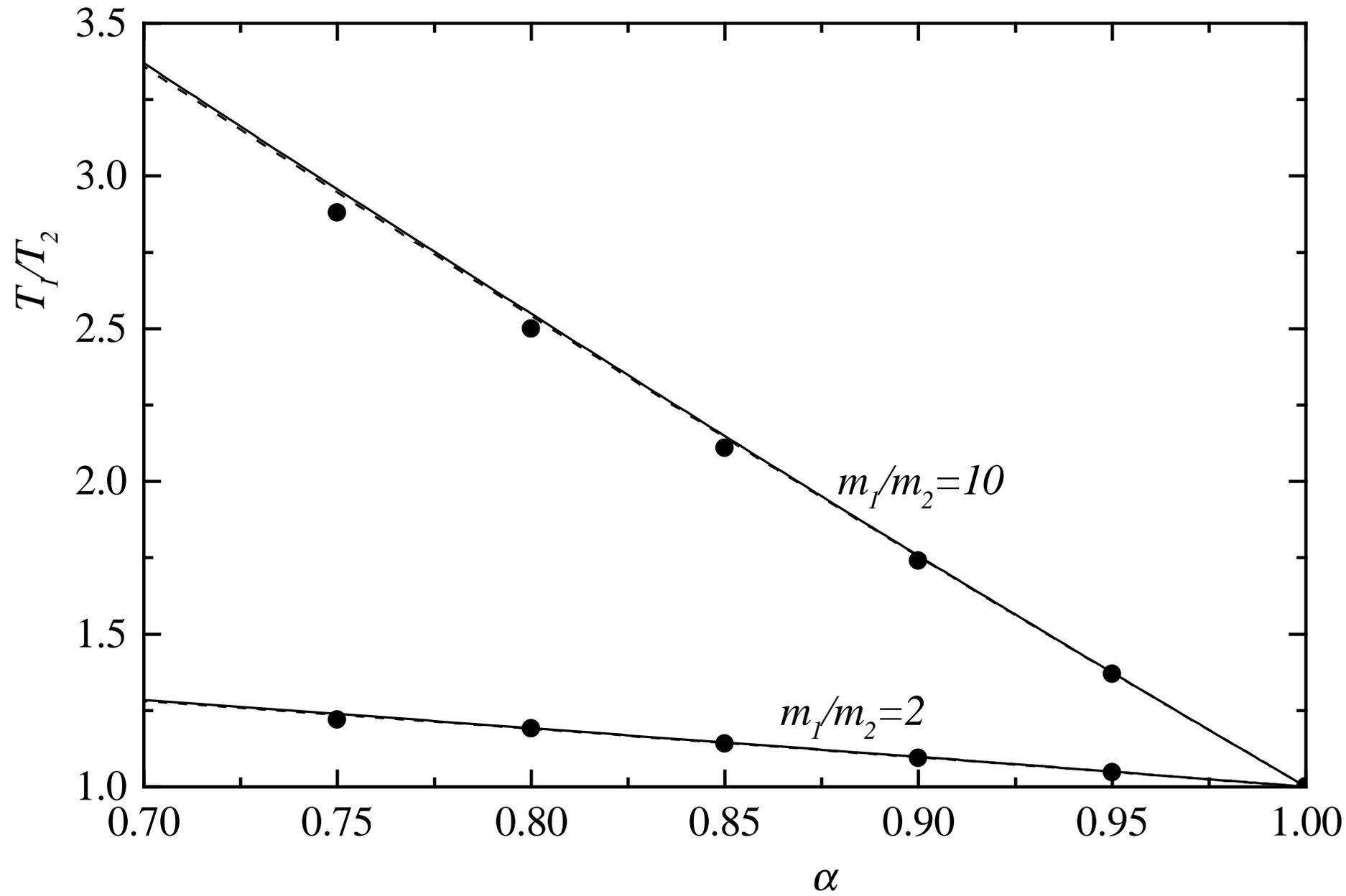
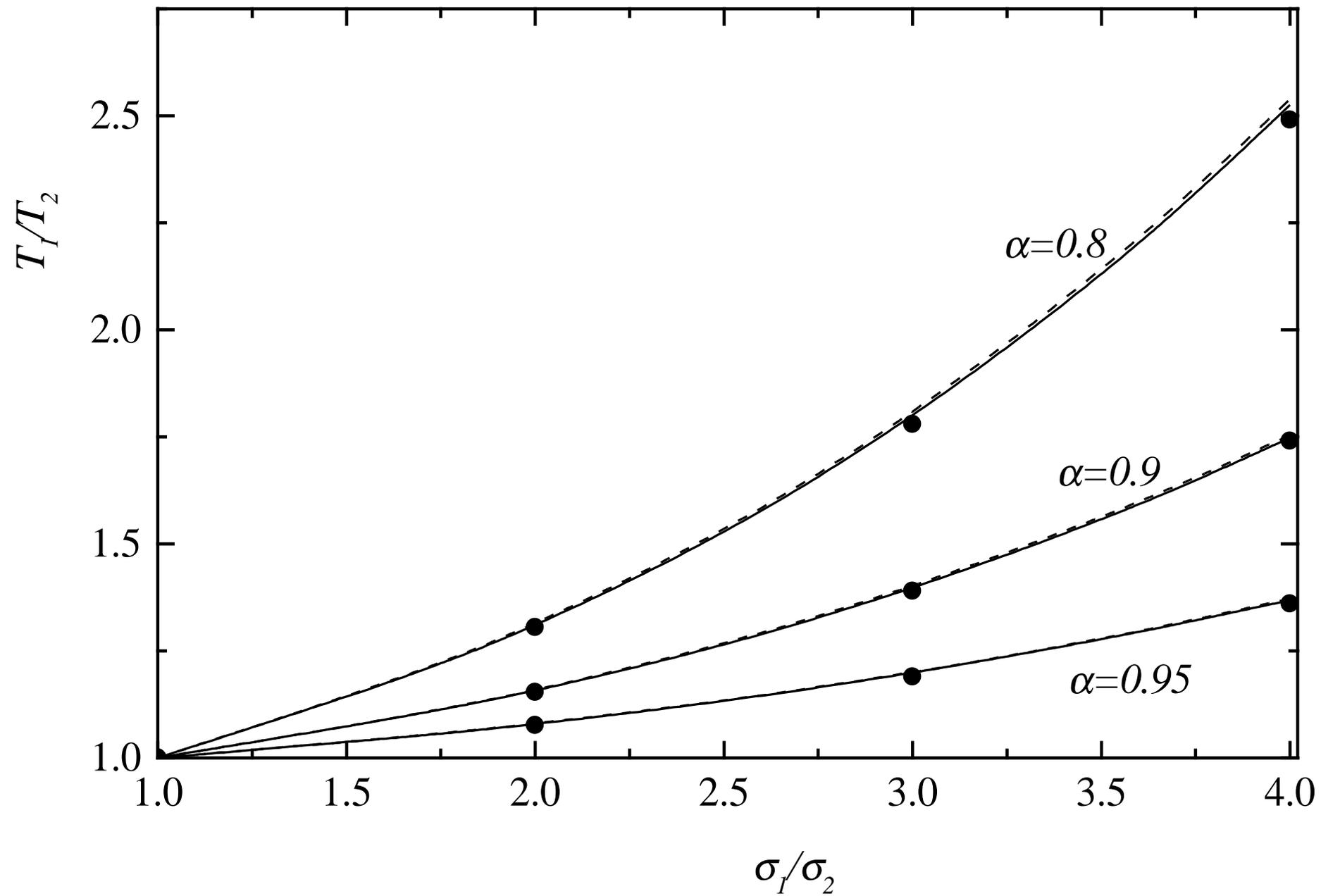
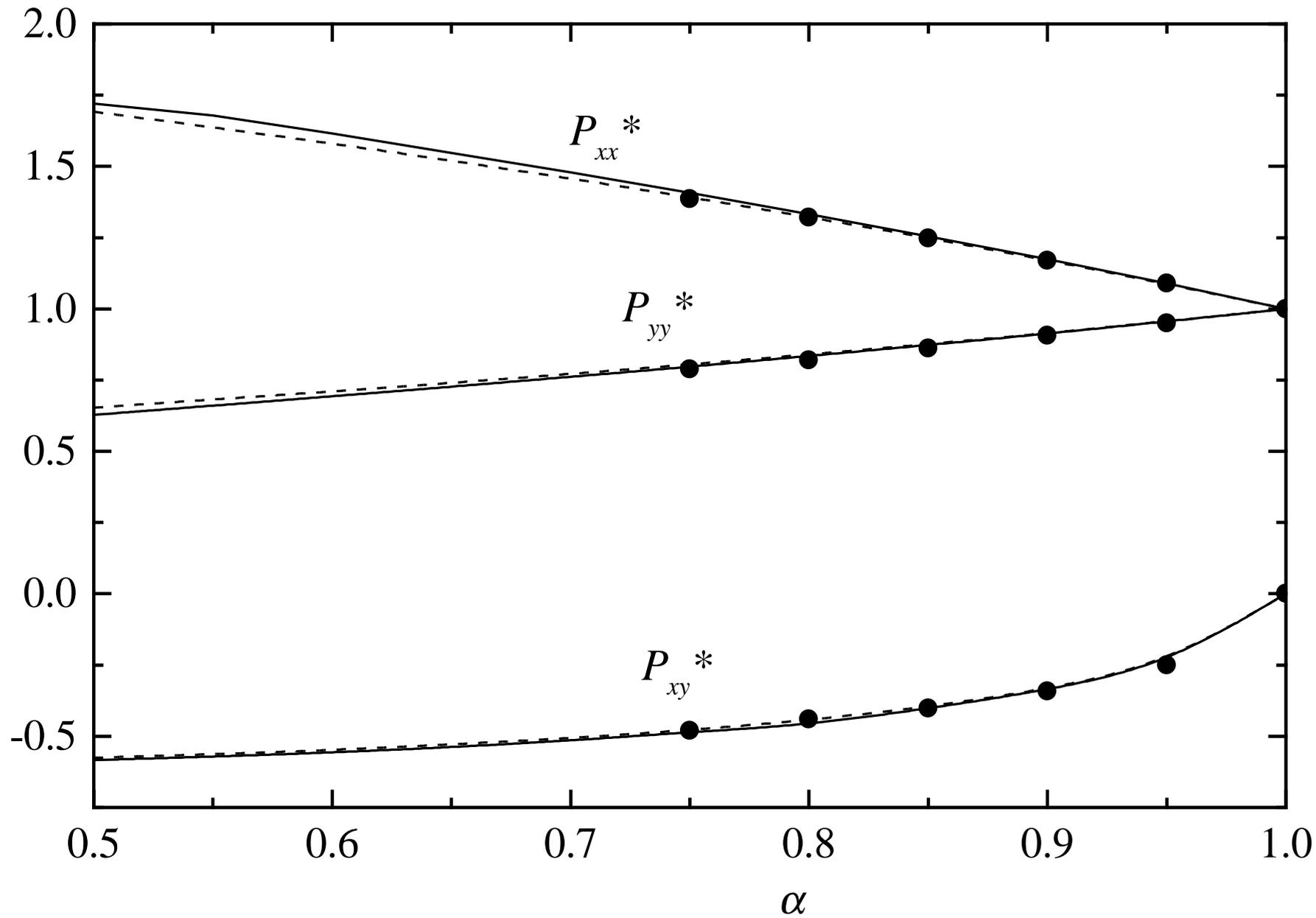
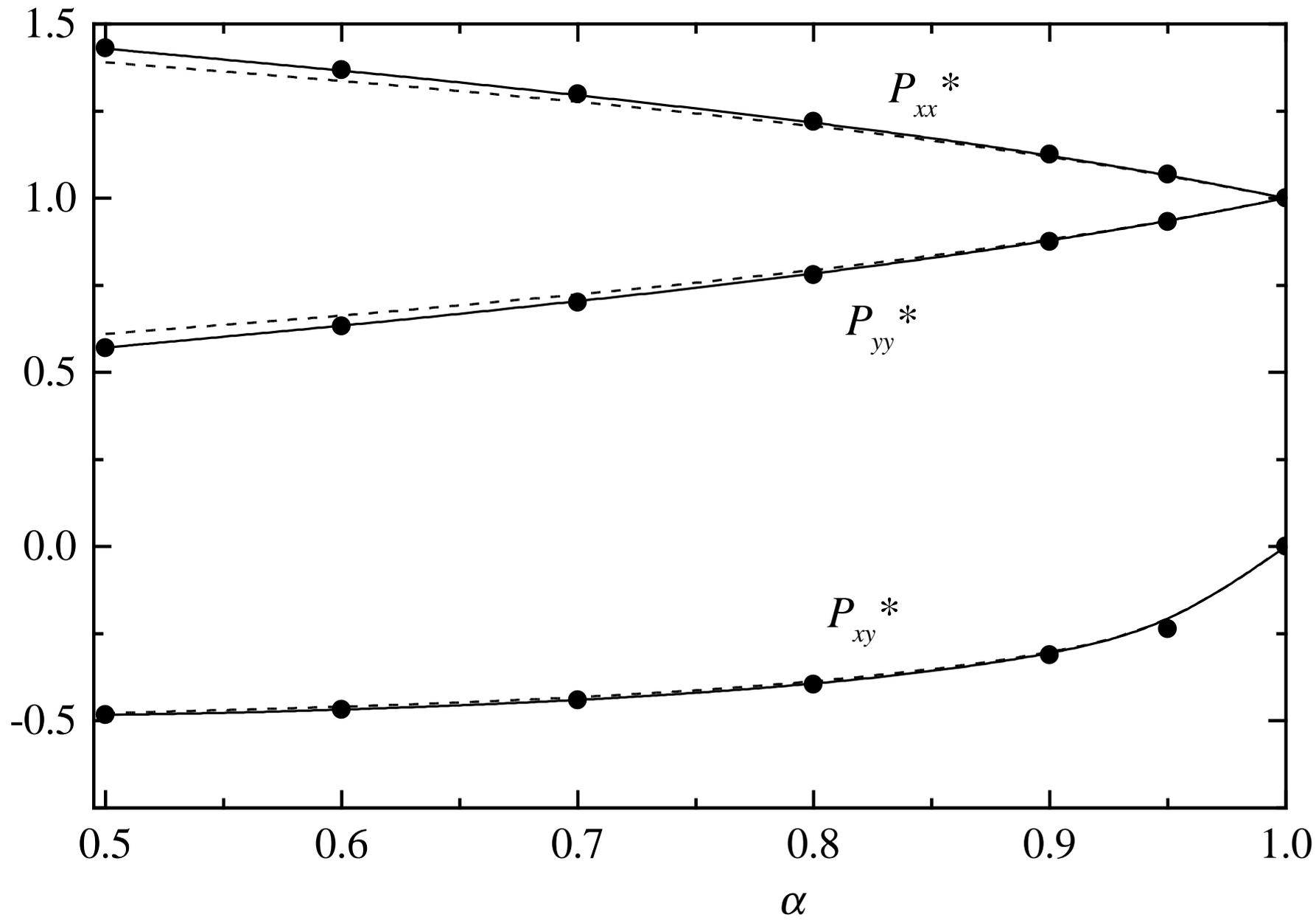
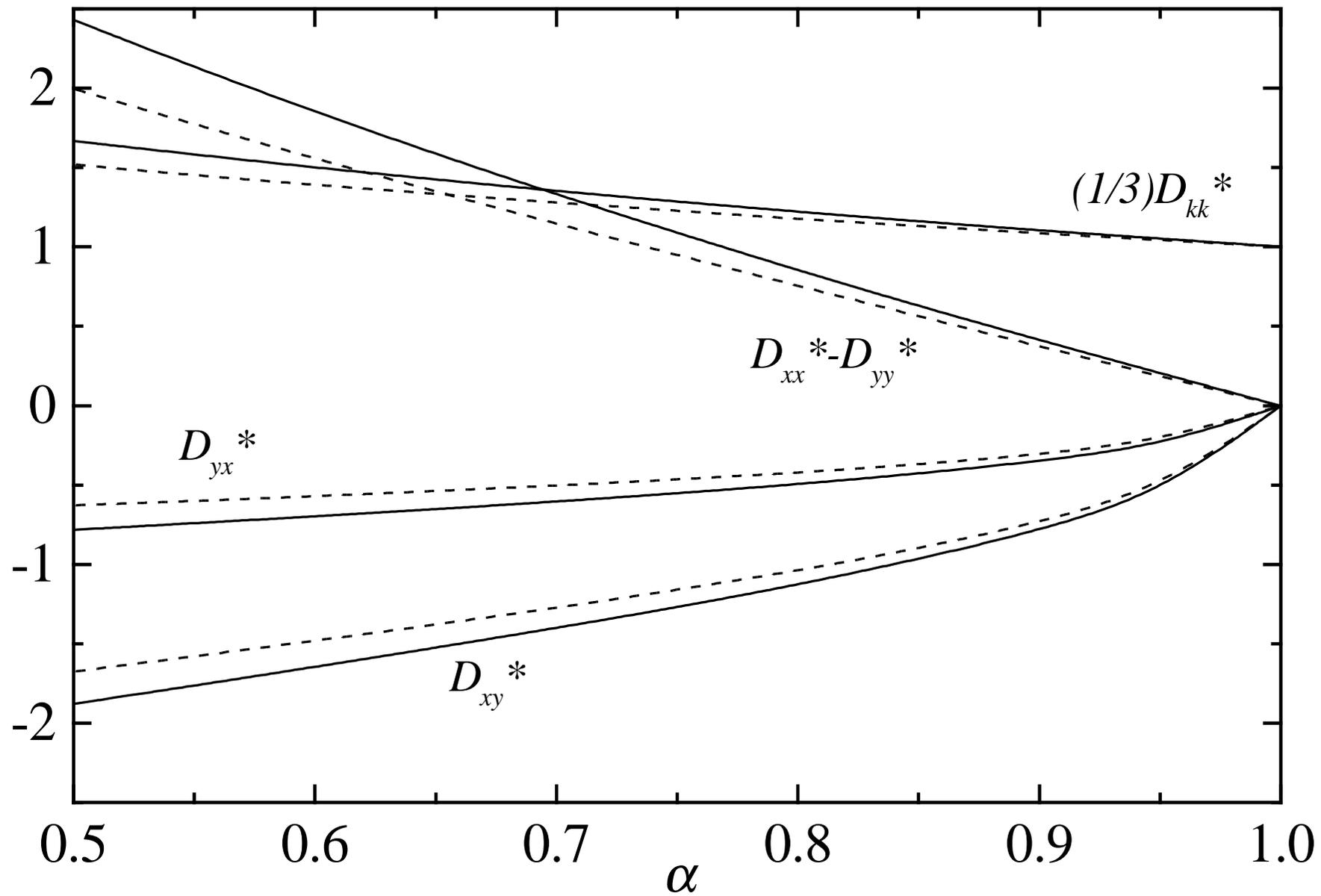
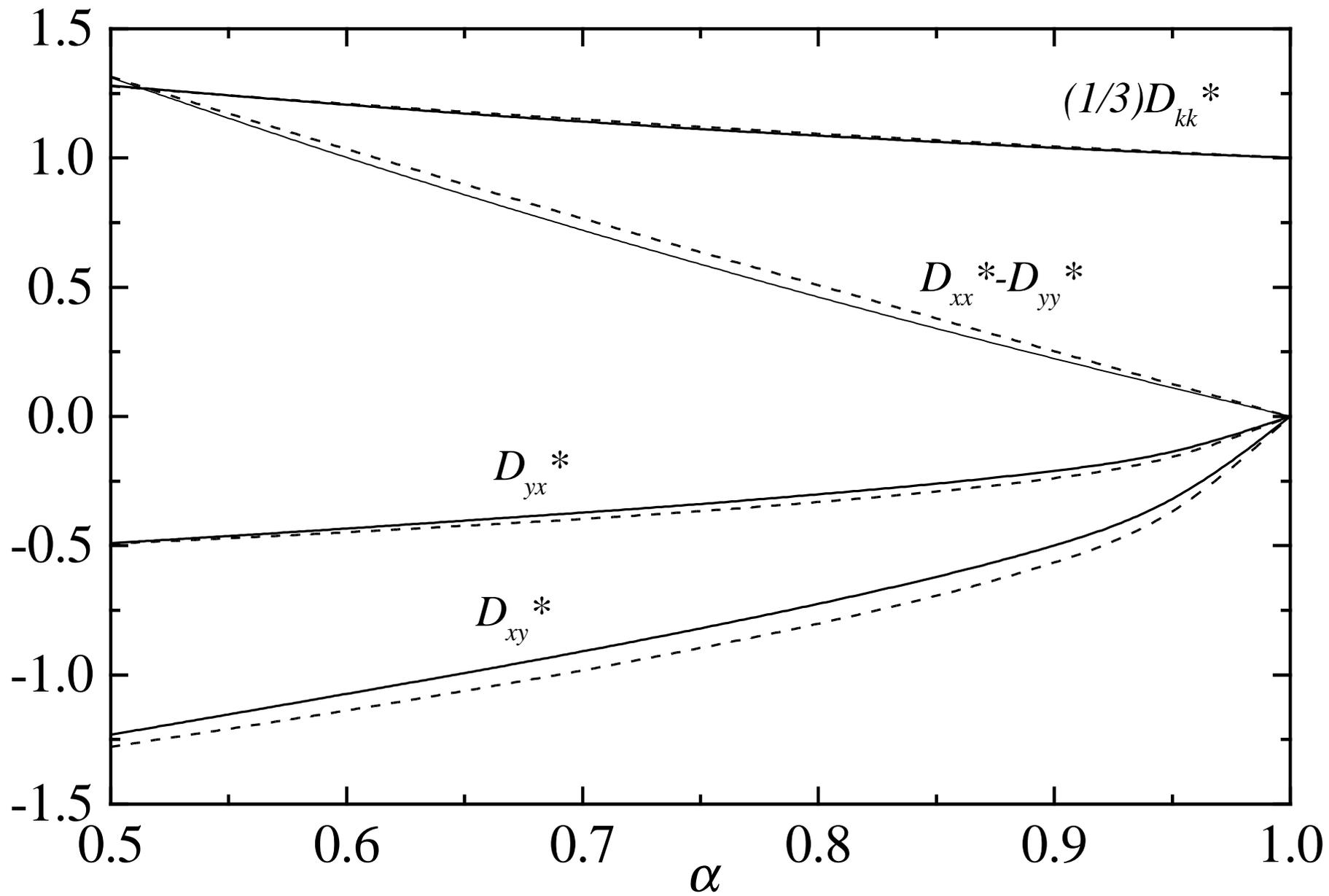
\end{document}